# Laser detection using liquid crystal polarization modulators

David M Benton[*]
Aston Institute of Photonic Technologies, Aston University, Birmingham, UK. B4 7ET

**Abstract**.
Lasers can be identified by their relatively long coherence lengths using interferometry. A Mach Zehnder interferometer incorporating liquid crystal polarization modulators is demonstrated as a means of low cost, robust laser detection. Temporal modulations, as a signature of coherence can be induced by modulating polarization changes in liquid crystal modulators, using low voltages. Sensitivities of less than 10nW can be achieved. The suitability as a means of laser detection is discussed.

**\*** E-mail: d.benton@aston.ac.uk

## 1 Introduction

Lasers are now a pervasive technology with many familiar applications that range from communication, material processing, 3D scanning, printing, medical applications and many more. The generation of laser radiation requires such an imbalance of thermodynamic equilibrium that only highly unusual astrophysical scale phenomena can generate lasing in nature [1]. All other lasers are engineered and used with a purpose. It is the intent behind the use that drives military desire to detect lasers. Military applications of lasers include range finding, target designation, laser dazzle and missile control [2]. The majority of lasers of interest to the military are pulsed lasers. The temporal resolution and high instantaneous brightness make them well-suited for operating over ranges of many kilometers in applications such as range finders and target markers. Generations of laser warner receivers (LWRs) have been developed to detect the threat posed by these lasers [3][4] and allow irradiated platforms to initiate appropriate countermeasures determined by the perceived threat. In the last few years a more widespread hazard has arisen from the preponderance of continuous wave (CW) visible laser diodes - so called laser pointers. Handheld visible lasers with powers of several Watts are available for a few hundred dollars with an ever-expanding range of wavelengths. These lasers have proven to be a menace when directed towards aircraft as they approach a landing, with over 1500 reports of lasers dazzling pilots in the UK and US last year. Conventional LWRs do not perform well at detecting these lasers because they rely on the rapid temporal changes in brightness observed with pulsed lasers. CW lasers are therefore a specific challenge in the world of laser detection.

Wang [5] divided laser detection into three categories - coherent recognition, scattering recognition and spectrum recognition. These are categories based on what is observed rather than the discriminating characteristic. Benton [6] took a classification approach based on discrimination technique - imaging, spectral and coherence. The first two categories are both essentially discriminating based on brightness. Imaging systems make use of CCD arrays [7][8] and intensified cameras[2], whilst semi-imaging systems look for tell-tale circles from bright sources[9]. Spectroscopic systems typically use dispersion with a diffraction grating and a detector array [10][11] Discrimination based on coherence detection is the main subject of this paper.

The concept of using the coherence properties of laser radiation as its signature of detection has been around for decades with the underlying principle being that broadband background radiation has a very short coherence length (typically microns) which suppresses interference effects in interferometers which are unbalanced (asymmetric). Fabry Perot interferometers with angle scanning have been used by Crane [12][13] for detecting pulsed and CW lasers. Michelson interferometers have been used as coherence detectors where long path length differences are able to determine the coherence length of the source [14][15] When used with an imaging system this is able to remove background illumination [16]. Multiple interferometer paths are able to act as a coherence length bandpass filter [17].

The Fizeau interferometer has been also been considered [18]. Benton [19] made use of a Mach Zehnder interferometer (MZI) to produce a low-cost laser detection system capable of discriminating wavelength. Cost is an important matter as typically interferometric detection systems can be fragile and (consequentially) heavy and expensive. This provides an economic asymmetry where the detectors for lasers are significantly more expensive than the laser threats they are detecting. Cost effective detectors will rebalance this asymmetry to some extent.

The requirement for robust detectors suggests that any system using a moving mirror may be overly fragile and thus a system with no 'moving parts' would be desirable. To meet this requirement an interferometric system based on polarization modulation has been investigated. Such a system has been investigated by Cohen [20] but the use of electrooptic modulators which were (at that time) bulky, require high voltage and are expensive do not meet the desire for low cost devices. In this paper polarization modulation is achieved through the use of liquid crystal modulators (LCM) which are compact, require low voltage and are relatively cheap to produce. Unlike the system of Cohen[20] the LCMs are incorporated into a MZI building on previous work [19] as a method of obtaining a low cost laser detection system.

**Methods**

A generic MZI with a polarization modulator is shown generically in fig.1.

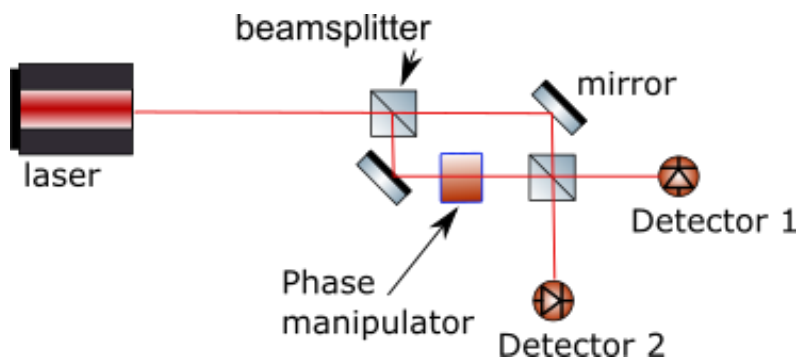

**Figure 1 A generic Mach Zehnder interferometer**

Introducing a polarization dependent component into a MZI causes 2 orthogonal polarizations to be present with independent interference regimes, but these are not distinguished by the detectors. Any interference seen by the detector is the net result of the 2 orthogonal polarizations detected simultaneously. (Note: using polarization sensitive beam splitters does not work here). Changing the applied voltage to the LCM modulates the refractive index of one polarization axis. Thus the time dependent intensity varies for only one polarization component. Laser sources are, in general, polarized and this will be at some unknown angle. By considering the simple but instructive case of linear input polarization we can model the expected system behavior.

The intensity in detectors 1 and 2, in each polarization mode (labeled x and y) can be given from the standard MZI interference equation [19][20]  as:

$$I_{1,x} = 2I_0 \sin(k\theta)^2 B^2 \left(1 + \cos\left(\pi \frac{v(t)}{V_\pi}\right)\right)$$

$$I_{1,y} = 2I_0 \cos(k\theta)^2 B^2$$

$$I_{2,x} = 2I_0 \sin(k\theta)^2 B^2 \left(1 - \cos\left(\pi \frac{v(t)}{V_\pi}\right)\right)$$

$$I_{2,y} = 2I_0 \cos(k\theta)^2 B^2 \quad (1)$$

Where $k = 2\pi / \lambda$, with $\lambda$ the wavelength, $\theta$ is the polarization angle of the incoming light relative to the x axis, $B$ is the beam splitting ratio assuming equal reflection and transmission (B=0.5) , $v(t)$ is the time varying voltage applied to the electrodes of the LCM and $V_\pi$ is the  voltage required to generate a phase shift of π. Plots of the normalized intensity level at one of the detectors can be produced as the input polarization is varied. The plots show the intensity at the minimum voltage level $V_0$ and the maximum voltage level $V_0 + V_\pi$. In each case the modulation amplitude is chosen to be $V_\pi$ as this gives the maximum modulation response in the detectors (although is wavelength dependent) . These plots can be seen in figure 2.  The top left plot shows that there are polarization values where the intensity at max and min values are the same and no modulation signal will be observed. This is because when the input polarization is orthogonal to the LCM modulation axis there is no effect. To overcome this we consider the case of 2 such LCMs, one in each arm, oriented orthogonally to each other.

$$I_{1,x} = 2I_0 \sin(k\theta)^2 B^2 \left(1 + \cos\left(\pi \frac{v_a(t)}{V_\pi}\right)\right)$$

$$I_{1,y} = 2I_0 \cos(k\theta)^2 B^2 \left(1 + \cos\left(\pi \frac{v_b(t)}{V_\pi}\right)\right)$$

$$I_{2,x} = 2I_0 \sin(k\theta)^2 B^2 \left(1 - \cos\left(\pi \frac{v_a(t)}{V_\pi}\right)\right)$$

$$I_{2,y} = 2I_0 \cos(k\theta)^2 B^2 \left(1 - \cos\left(\pi \frac{v_b(t)}{V_\pi}\right)\right)$$

$$(2)$$

Here we label the voltages with the subscripts *a* and *b* to represent the modulation voltages applied to each modulator. Applying the same signal to both modulators (oriented orthogonally) we get the result for min and max intensity values as shown in the top right of figure 2. Here we can see that there is a modulation signal to be observed at all polarizations. By adding a relative phase difference to the modulators we can change the behavior as shown with a π/2 phase difference (bottom right) and a π phase difference (bottom left).  The amplitude of the modulating signal in the detector is the difference between the values for the max and min applied voltages. To represent the signal size obtained from a balanced system we must subtract signals from the 2 detectors, which are in antiphase. Figure 3 shows plots of the system signal strength for differing polarization values where the phase difference between the signals to each modulator are given as a fraction, *f*, of $\pi$ . It would appear that a system with no phase difference between modulators would be most reliable but it is possible that the behavior seen with some phase difference could be useful in helping to determine the incoming wavelength, such as by observing the difference in signal as the phase is changed.

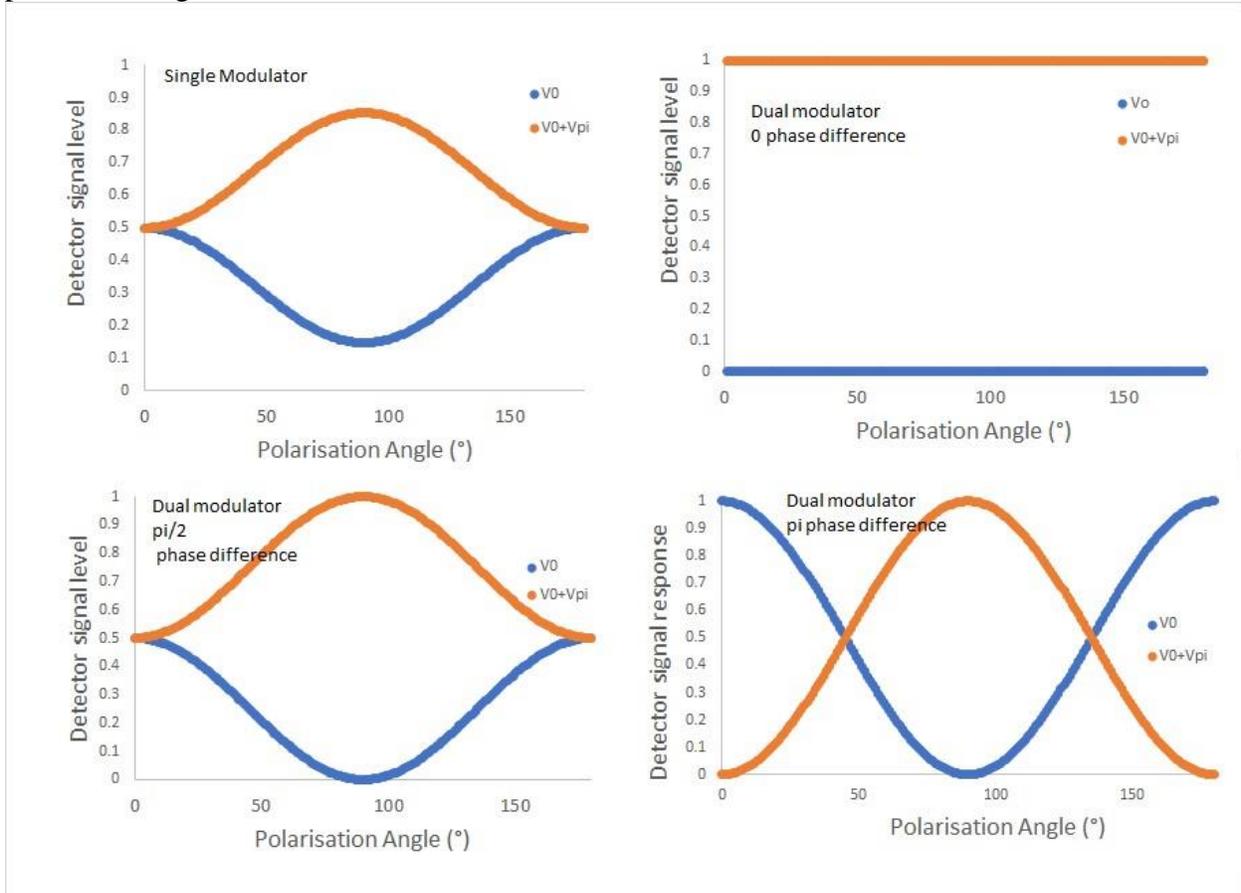

**Figure 2. Plots of modelling of signal intensity at one of the detectors as input polarization is varied. All plots show the normalized intensity at the min ($V_0$) and max ($V_0+V_\pi$) voltages. Top left is for a single modulator. Top right is for 2 orthogonal modulators with no phase difference between signals. Bottom left has a $\pi/2$ phase difference. Bottom right a $\pi$ phase difference.**

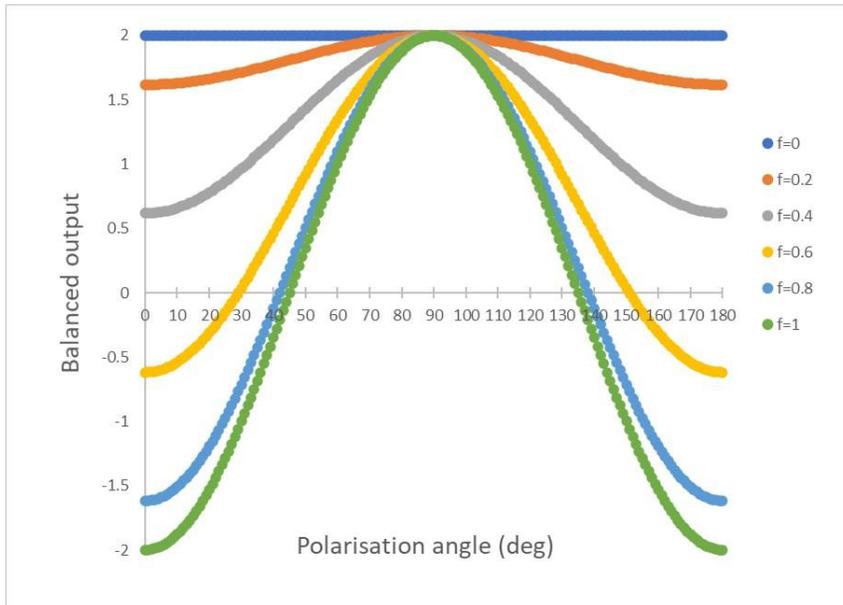

**Figure 3. balanced detector relative signals against polarization angle, for differing phase offsets between modulators.**

**Experimental setup**

A MZI incorporating LCMs is shown schematically in Figure 4. This system is composed of 2 non polarizing beam splitters dividing light towards 2 adjustable mirrors. Two polarization modulators (Thorlabs LCC1111U-A) - were located one in each path. The electrical signals from 2 silicon photodiodes were amplified and sent to a data acquisition unit (National Instruments USB 6341) and then to a computer where the signals were processed using a LabView program.

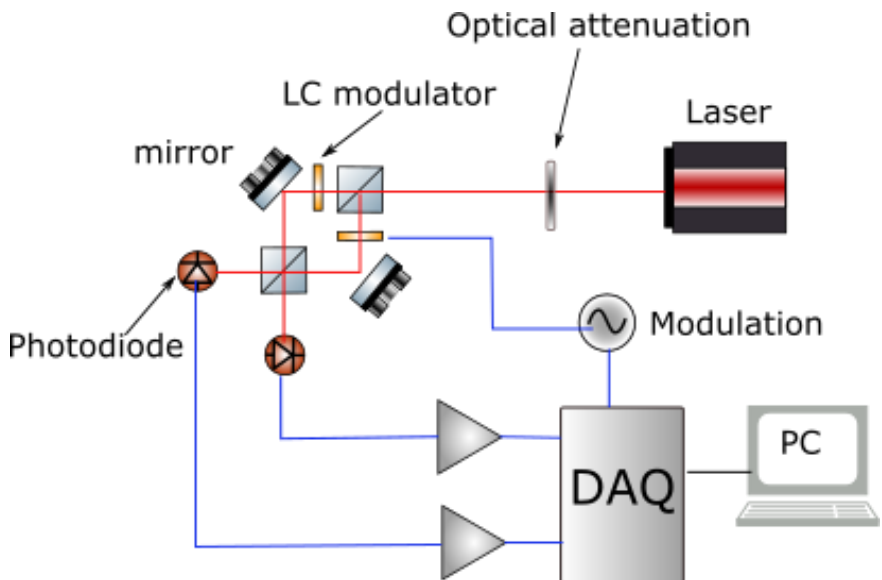

**Figure 4. A schematic diagram of the laser detection system with polarization modulators in a MZI**

Applying a voltage across the LCM requires the regular polarity reversal of the voltage across the device electrodes in order to prevent the permanent displacement of the LC medium causing performance reduction. This polarity reversal is usually done at a frequency of a few kHz. Modulating the device voltage to change polarization properties involves changing the amplitude of this "carrier" frequency. A modulation waveform was generated using the DAQ device with 2 analog to digital converters connected to the LCM electrodes. The amplitude, carrier frequency and modulation frequency were controlled using a LabView program.

**Results**

It is first necessary to characterize the response of the LCM. This was done using a polarizing beamsplitter with 2 photodiodes measuring the intensity of each output polarization, and an input laser polarized at 45°. A modulation frequency was applied to the LCM and the resulting intensities at the detectors were subtracted to produce a measure of the amplitude of the polarization effect being induced. The laser wavelength being used was 635nm from a laser diode and the amplitude of the applied voltage was +/- 2.5V which had been found to produce a significant level of response for this wavelength. A plot of the polarization response vs modulation frequency is shown in Figure 5. This clearly shows that the amount of polarization response drops rapidly as the frequency is increased. The trendline is a 4th power polynomial. Thus it can be seen that it is preferential to operate these devices at relatively low frequencies of a few 10s of Hz.

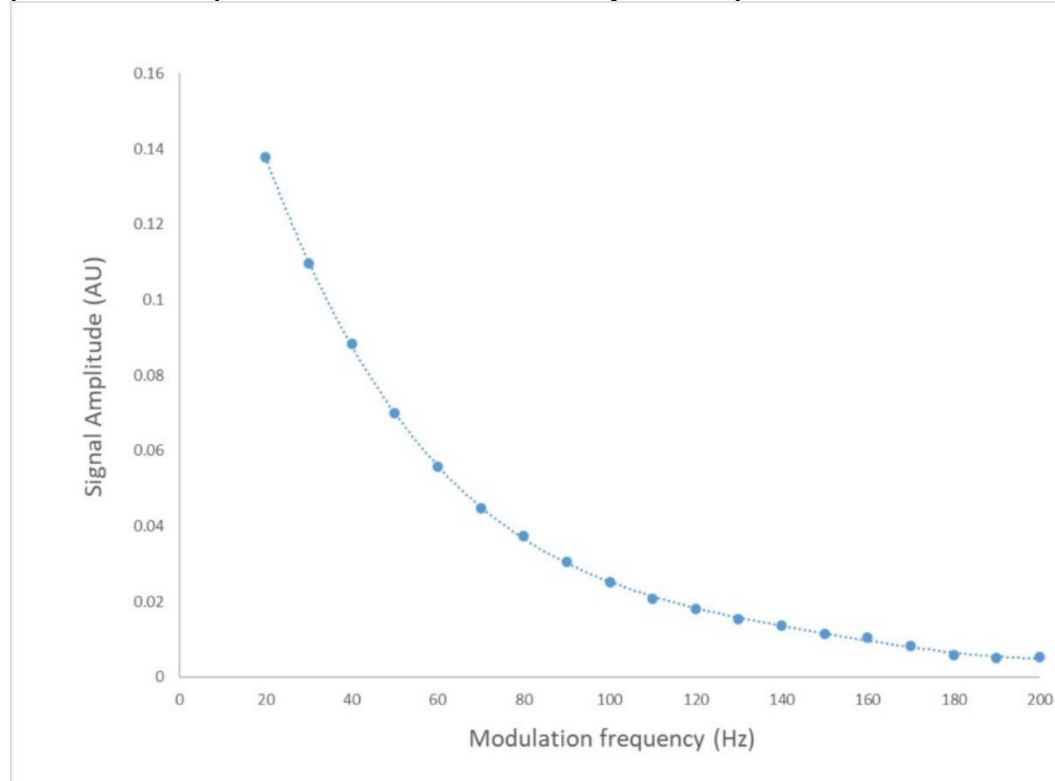

**Figure 5. A plot of the polarization effect produced by the LCMs as frequency is varied**

Using the MZI setup a shown in Figure 6, the interferogram in fig 4. shows the signal in both detectors for an applied modulation frequency of 32Hz with a sampling rate of 20kHz and an input laser at 630nm. This clearly shows the modulation of the signal in both detectors changing in antiphase.

As was the case in [19] the strength of the received signal at the modulation frequency is measured by taking the Fourier transform of the interferogram and determining the power at the modulation frequency.

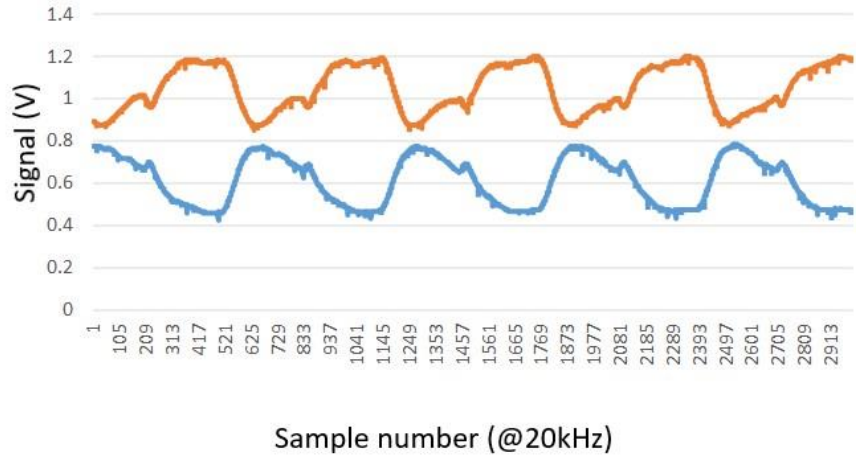

**Figure 6. Modulated intensity seen in both detectors as a modulating voltage is applied to a LCM**

Varying the amplitude of the modulation alters the amount of path length difference for one polarization, an effect which is wavelength dependent. Figure 7 plots the detected signal power against the amplitude of the modulation applied (at 20Hz) to the LCM. This is shown for two wavelengths - 632nm and 532nm. This was also attempted at 405nm but no modulation could be observed, despite the device specifications claiming operation at 350nm. An amplitude of around +/-2.5V is a useful level giving good response.

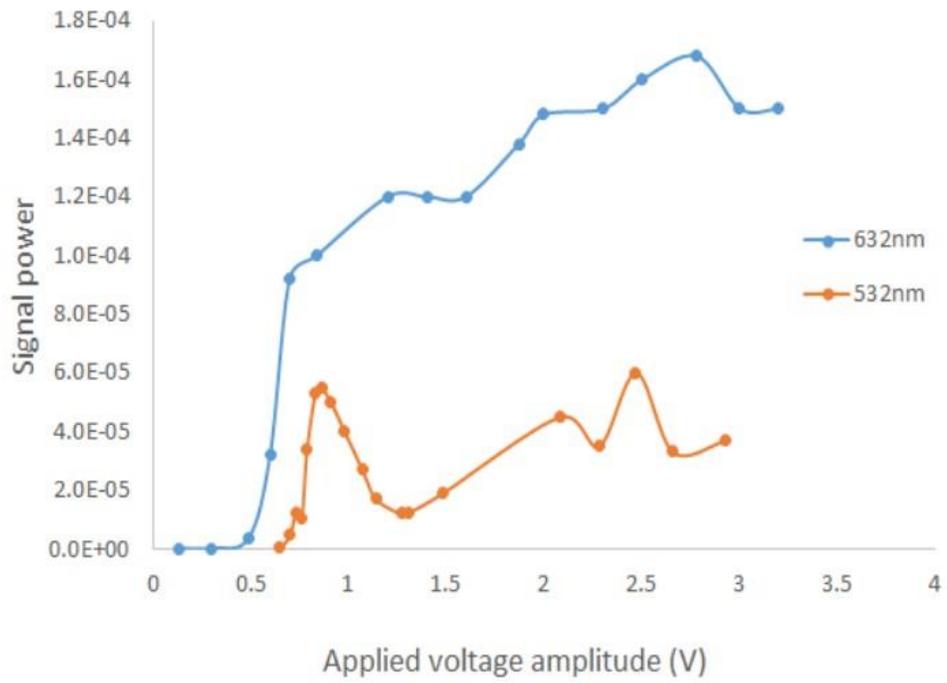

**Figure 7. Signal power at the modulation frequency for varying amplitude of drive modulation**

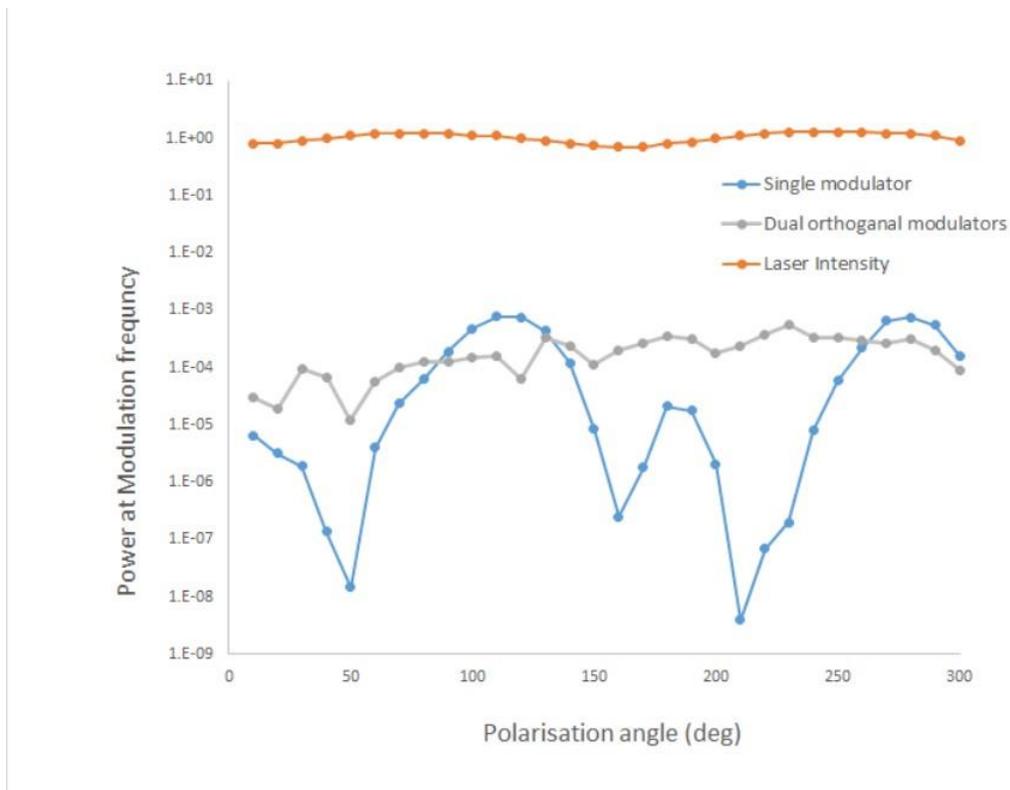

**Figure 8. The variation of MZI system response to changing in put polarization. Data shows the laser intensity, the effect for a single modulator and a dual modulator system.**

The effect of the polarization angle of the input laser was examined and can be seen in Figure 8. The system was driven at a frequency of 20Hz and with an amplitude of 2.5V. The DC level of the detected signal is shown representing the laser power and showing that there is only a small variation of power with polarization angle. With a single modulator operating there were found to be two positions with maximum response and two with very low response, corresponding to polarization alignment with the LCM axes.

Clearly this means that there are situations where even an intense laser would generate no modulation power at certain polarization angles. Such a device would have only limited utility. The process was repeated with a second LCM oriented orthogonally to the first but driven with the same modulation. This shows that modulation power is observed at all input polarization angles. Using two LCMs presents us with the possibility of controlling the devices independently. In the present case this caused a problem as the DAQ device has only two analog outputs. However the convenient choice of +/- 2.5V applied to each electrode is equivalent to switching between 0V and 5V, and hence 4 digital output lines could be used. Whilst not allowing independent control of amplitude, the LCMs can be modulated at different frequencies which could enable, for example, looking for the resultant modulation power at the sum frequency. which may have advantages such as better noise. The digital system switched at a carrier frequency of 2kHz with the modulation frequency applied by shifting the phase of the two electrodes by pi every half period. Two independent frequencies were observed in the detector output but at this point it was clear that one of LCMs had significantly degraded (it was older and had more use) and produced less modulation power. Also the net effect was to reduce the modulation power relative to both LCMs having the same frequency. Whilst no significant results can be presented for this it is included for completeness.

The detection sensitivity of the system was examined using the digital modulation with both LCMs present. This was conducted with a diode laser at 635nm and various levels of attenuation using neutral density filters. Consecutive Fourier transform power spectra were summed to increase sensitivity by increasing integration time. Various modulation frequencies and integration times were examined. Results of the signal to noise levels recorded are shown in Figure 9. As expected sensitivity at the highest frequency examined (72Hz) is worst with a sensitivity limit (S:N=1) of around 20nW (5s integration). The best response for a 20Hz modulation signal with a 10s integration time suggest that a sensitivity of around 1nW is possible. Normalising the responses to a 1second integration time suggest a detection threshold of 10nW at 20Hz and 60nW at 72Hz.

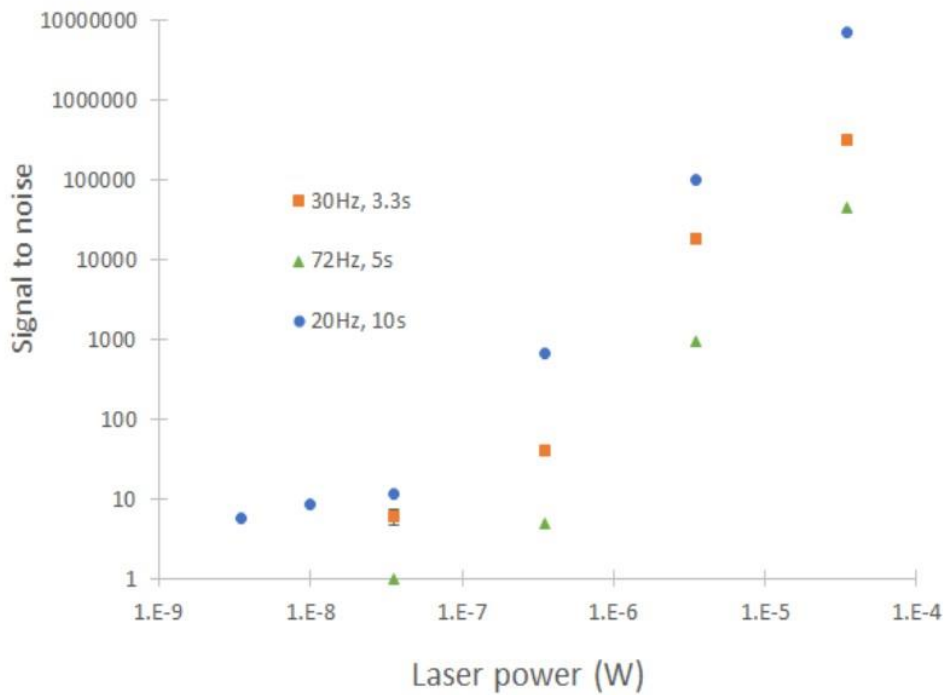

**Figure 9. Plots of the signal to noise for different modulation drive frequencies and differing signal collection times.**

## Conclusions

The motivation behind this work was to continue the development of low cost laser detection as given in [19] based upon coherence detection. The use of LCMs replaces the piezo modulated mirror as it is always desirable to have no moving parts, particularly those with critical alignment. The LCMs have been successfully used within an asymmetric MZI as a means of modulating coherent light and with no moving parts they certainly lead to a more robust system that is easier to align and maintain. The use of two LCMs oriented orthogonally has been used to overcome the limitation of response to particular polarization angles seen with a single modulator. The LCMs require only low voltage signals which is an advantage over say electrooptic polarization modulators and have been used with a digital drive signal. Their use has been targeted at CW lasers, in particular the laser pointer type. These lasers may typically have short coherence lengths (<1cm, sometimes <1mm). The asymmetry of the MZI arms needs only to be a few microns of length difference to prevent background contributions, which is easily achievable. Thus this is a viable technology for laser pointer detection.

However, there is no hiding the limitations that arise from the LCMs themselves. Most notably the limitation of the inherent switching frequency of the devices. The LCM response is best at low frequencies around a few 10s of Hz which naturally limits the speed and sensitivity of response of the system. In critical situations this slow response would be unacceptable. Also, the effect of atmospheric scintillation leading to intensity modulations upon a transiting laser are significant at these low frequencies, especially in comparison to devices that operate beyond 1kHz. The LCMs chosen were designed for use in the visible part of the spectrum. LCMs are available for the near IR but these devices have a thicker layer of liquid crystal and have a slower response time.  However the slow modulation rates of LCMs are very well suited to use with cameras instead of photodiodes. The modulation rates can show up temporal modulations with spatial locations defining the origin of the laser source with video processing. This may be a more promising line of future investigation.  As with most technologies there are particular situations for which they are well suited. These LCM based interferometers have a slower response and are less sensitive which limits their applicability, but may be relevant when faster LCMs or alternative devices are available.

## *References*

[1] Chen S. "Alien Light" https://spie.org/news/photonics-focus/janfeb-2020/astrophysical-lasers?SSO=1

[2] Dubois, J., & Reid, F. "Detecting laser sources on the battlefield". Proc. SPIE, 6796(2007), 67962F (2007). http://doi.org/10.1117/12.779234

[3]  Pietrzak, J. "Laser warning receivers". Proc. SPIE, 5229(2), 318–322. (2003)

[4] Dąbrowski, M., Młodzianko, A., Pietrzak, J., Zygmunt, M., & Niedzielski, R. "Laser warner receiver LWR-H". Proc SPIE, 6598, 65980S–4(2006). http://doi.org/10.1117/12.726581

[5] Wang, L. et al. "Optimum design of wideangle laser detecting system based on fisheye lens and sinusoidal amplitude grating". Optics Communications, 310, 173-178 (2014).

[6] Benton, D. M., Zandi, M. A., & Sugden, K. (2019, October). Laser detection utilizing coherence. In Technologies for Optical Countermeasures XVI (Vol. 11161, p. 111610G). International Society for Optics and Photonics

[7] Kumar, S. et al. "Design of a laser-warning system using an array of discrete photodiodes— part I". Journal of battlefield technology 14.1 (2011).

[8] Ying, J. and Zhou, Z. "Study on Image Processing Technology in Imaging Laser Detection System". IEEE (2010).

[9] Tipper, S., Burgess, C., & Westgate, C. (2019, May). Novel low-cost camera-based continuous wave laser detection. In Situation Awareness in Degraded Environments 2019 (Vol. 11019, p. 110190B). International Society for Optics and Photonics

[10] McAulay, A. D. "Detecting modulated lasers in the battlefield and determining their direction". Proc. SPIE, 7336, 73361J. (2009). https://doi.org/10.1117/12.819423

[11] Zhang, J., Tian, E., & Wang, Z. "Research on laser warning receiver based on sinusoidal transmission grating and high speed DSPs". WSEAS Transactions on Circuits and Systems, 5(8), 1366–1371. (2006). https://doi.org/10.1088/1742-6596/48/1/152

[12] Crane Jr, R. "Laser detection by coherence discrimination". Optical Engineering, 18(2), 182212. (1979).

[13] Crane Jr, R." The angle-scanned interferometer". Optical Engineering, 18(2), 182205 (1979)

[14] Hickman,D. "An optical sensor based on temporal coherence properties," J. Sci. Instrum. 21, 187–192 (1988).

[15] Coutinho, R. C., French, H. A., Selviah, D. R., Wickramasinghe, D., & Griffiths, H. D. "Detection of coherent light in an incoherent background [for IRST]". In 1999 IEEE LEOS Annual Meeting Conference Proceedings. LEOS'99. 12th Annual Meeting. IEEE Lasers and Electro-Optics Society 1999 Annual Meeting (Cat. No. 99CH37009) (Vol. 1, pp. 247-248). IEEE. (1999, November)

[16] Duffey, C. J. and Hickman, D. "An imaging system based on temporal coherence differences," J. Phys. D: Appl. Phys. 21, S56–S58 (1988).

[17] Sutton, P. "A novel electro-optical remote-sensing technique based on bandpass coherence processing". J. Phys. 22 (1989).

[18] Russell J.C. "Coherent laser warning system". Pat. 6,151,114. 2000.

[19] Benton, D.M. "Low-cost detection of lasers". Opt. Eng. 56(11), 114104 (2017).

[20] Cohen, J. D. "Electrooptic detector of temporally coherent radiation". Applied Optics, 30(7), 874–883. (1991). https://doi.org/10.1364/AO.30.000874

**Biography**

David Benton graduated in Physics from the University of Birmingham in 1989. He completed a PhD in laser spectroscopy for nuclear physics in 1994 and then conducted postdoctoral research in positron emission tomography and then laser spectroscopy for nuclear physics, all at the University of Birmingham. In 1998 he joined DERA which became QinetiQ where he worked on a variety of optical projects. He was the leader of a group building quantum cryptography systems and was involved in a notable 140km demonstration in the Canary Islands. He became Chief Scientist for L-3 TRL in 2010 working on photonic processing techniques for RF applications. He is now at Aston University with a variety of interests including novel encoding techniques, gas sensing and laser detection techniques. He is a member of the SPIE.